\documentclass[11pt,twoside]{article}
\usepackage{ap-article}

\usepackage{graphicx}
\usepackage{graphics}
\usepackage{psfig}
\usepackage{xcolor,stfloats}
\usepackage{url}
\newenvironment{figurehere}
{\def\@captype{figure}}
{}
\def\labart{yourLabel}      
\shortauthors{Shengli Li et al.}
\shorttitle{Research of the Correlation between the H1N1 Morbidity Data and Google Trends in Egypt}
\title{Research of the Correlation between the H1N1 Morbidity Data and Google Trends in Egypt}
\author{%
Shengli Li\,\up{1}\,\,,
Xichuan Zhou\,\up{1}\footnote{Corresponding Author: zxc@cqu.edu.cn\\
This research was supported in part by the national natural science foundation of China (contract 61103212, 61471073), and by the post-doctoral science foundation (contract 2012M521678, 2013T60836 and XM20120036)} \,,
Qin Li\,\up{2} \,,
Han Zhao\,\up{2}
}
\addresses{%
\up{1}
College of Communication Engineering, Chongqing University\\
Chongqing, 400044, China%
\\ \vspace*{0.04truein}
E-mail: lishengli1024@163.com, zxc@cqu.edu.cn
\\ \vspace*{0.05truein}
\up{2}
the Centers for Disease Control and Prevention of Chongqing.\\
Chongqing,400042, China
\\ \vspace*{0.04truein}
E-mail: liqin@vip.163.com, zhaohan@vip.163.com
}
\begin{document}
\label{\labart-FirstPage}

\maketitle
\abstracts{%
The search engine based on influenza monitoring system has been widely applied in many European and American countries. However, there are not any correlative researches reported for
African developing countries. Especially, the countries Egypt has not designed an influenza monitoring system on the basis of the Internet search data. This study  aims at analyzing the correlation between the Google search data and the H1N1 morbidity data of Egypt, and examining the feasibility of Google Flu Model in predicting the H1N1 influenza trend.
}
\par\bigskip\par
\keywords{Goolge Trends, Goolge Flu, H1N1 in Egypt, Correlation}

\vspace*{10pt}\textlineskip
\begin{multicols}{2}

\section{Introduction}
The frequent mutation of influenza virus in the past century caused several pandemic\up{1}, posing a great threat to the world. As the result, the prevention and prediction of the influenza have become a global public hygiene topic\up{2-4}. Traditional surveillance system plays an important role in the control and prediction of the contagion, but has one or two
weeks delay\up{5}. And even one week delay can bring significant difficulty in disease prevention and control. For instance, on April 27, 2009, only the United States, Canada and Mexico
reported H1N1 cases, but after two weeks, 30 countries reported H1N1-infections to the WHO\up{6-7}. By December 2013, the global epidemic of H1N1 spread to 214 countries and regions, and caused 15934 deaths\up{6}, which raised serious concerns in the society. So far, the WHO has monitored 130 countries and regions in the world for influenza outbreaks, but only covers 26 out of 59 countries and regions in Africa\up{7}. It would be a horrible disaster in such areas if epidemics outbreak. And it can work better in prevention if there is a real-time syndromic monitoring system for African countries\up{8,9}.

With the popularity of Internet, a growing number of people used search engine to seek things that they concerned. A new way of monitoring influenza epidemic, using the Internet search data, came to be known, which attracted a lot of researchers\up{4,10-13} and made people aware of the hope in real-time prediction of epidemics. Based on that, Google launched Google Flu Trends (GFT) to predict the influenza trends\up{14}. And this method can potentially provide surveillance results one or two weeks earlier than the traditional monitoring system\up{11-13,15-16}. So far, 29 countries have launched the GFT, which are mainly in the European and American countries\up{14}. And few research extended the GFT method to African
countries. Accordingly, we examined the correlation between the Google Trends data and the H1N1 morbidity data, and the feasibility of Google Flu Model in Egypt.

\section{Method}
It was a novel low-cost method to predict the influenza trends using the data from search engine in European and American countries, but few research focused in developing
countries\up{12-13,15-18}. This study focused in Egypt, and the examined period was from January 4, 2009 to December 31, 2013(261 weeks).

We used ``corrcoef'', a function of the Matlab software, to calculate the Pearson¡¯s correlation coefficient. We calculated the total and annual correlation coefficients between the H1N1
influenza surveillance data of Egypt and Google Trends data for the overall study period.

The changes of Pearson¡¯s correlation coefficient, when the H1N1 influenza surveillance data of  Egypt data were shifted ahead or behind the H1N1 morbidity data for one or two weeks, were also examined. In our experiments, ``lagging'' was defined as the H1N1 morbidity data were shifted behind the Google search data, and ``preceding'' was defined contrarily. Strong correlation was defined when Pearson¡¯s correlation coefficient was greater than 0.7.

Suppose for the \emph{t}th week, the Google search volume of the \emph{i}th search query was $x_{it}$ ($i$=1,2\ldots ,$n$ ) and the H1N1 cases number of Egypt of the  \emph{t}th week was set as variable $ y_{t}$. The Google Flu Model could be written as equation 1 \up{12,17}.
$$y_t=\beta_0+\beta_1*x_{1t}+\beta_2*x_{2t}+\ldots+\beta_n*x_{nt} \qquad (1) $$
where $n$ was the number of search queries. The least square method was applied to calculate the parameters $\beta_j$($j$=0,1,2\ldots ,$n$), confidence interval and significance probability. The parameters in the model were updated once a week. We also repeat the experiment when the H1N1 cases number of Egypt were shifted ahead or behind  the Google search data
for one or two weeks. Then the correlation coefficient  between the model estimates  and the actual H1N1 morbidity trend was calculated.

The H1N1 case number of Egypt were downloaded  from the WHO's website  as weekly data form\up{7}. These data were provided by the Global Influenza Surveillance Response System(GISRS)
and National Influenza Reference Laboratories collaborating with the  GISRS. The data were updated every other week\up{19}.

In Egypt, English is used in northern areas, but the official language is Arabic. First we examined the English search queries related to H1N1 from the previous related researches and the
Wikipedia (such as flu names, symptoms, treatments, prevention methods, etc.)\up{5,6,20}. Then, these search queries were translated into Arabic. By setting the location parameter to ``Egypt'' and the time parameter to ``2009-2013'', we downloaded the search volume data of the selected search queries as weekly data form from Google Trends. The downloaded search volume data were normalized between 0 to 100 by Google, to cancel out the variable's effect on the data.  When we searched these search queries, Google provided the ``related searches'' at the bottom of the website. We recommended these search queries provided by Google. Finally we achieved 30 English search queries  and 20 Arabic search queries which had 11 of 30 English search queries and 4 of 20 Arabic search queries selected from the previous related researches, respectively. There were 26 search queries had  a positive correlation coefficient compared with the H1N1 morbidity data of Egypt(Table 1). And the other 24 search queries' volume was close to zeros, they were ``bad cold'', ``bird flu'', ``cold and flu'', ``cough'', ``H1N1 symptoms influenza'', ``influenza A virus'', ``new flu'', ``new influenza'', ``sore throat'', ``swine flu H1N1'', ``swine flu treatment'', ``swine influenza'', ``zoonotic swine flu'', ``virus H1N1'', ``zoonotic swine flu'' in English, and ``bad cold'', ``bird flu'', ``H1N1 Egypt'', ``H1N1 vaccine'', ``H1N1 virus'', ``swine flu case'', ``swine flu vaccine'', ``virus H1N1'', ``zoonotic swine flu'' in Arabic.

\newcommand{\tabincell}[2]{\begin{tabular}{@{}#1@{}}#2\end{tabular}}
\vglue13pt
  \tcap{Search queries related with H1N1 influenza which were found in previous similar researches, the Wikipedia website and 'related searches' provided by Google.  The following search queries
  had a positive correlation coefficient compared with the H1N1 morbidity data of Egypt.}
  \vglue-6pt
  \centerline{\small\baselineskip=13pt
  \begin{tabular}{c c c c}\\
    \hline
    \textbf{English}&&\textbf{Arabic}&\\
    \hline
    H1N1&                      0.50&                search swine flu      &0.58\\
    H1N1 vaccine&	           0.43&                for swine flu         &0.55\\
    virus H1N1&                0.39&                Egypt swine flu       &0.53\\
    swine flu vaccine&         0.38&               \tabincell{c}{Symptoms of\\swine flu}&0.53\\
    Tamiflu&                   0.37&                Swine flu             &0.49\\
    swine flu symptoms&        0.33&                multi swine flu       &0.49\\
    swine flu&    	           0.30&                pig flu               &0.48\\
    H1N1 Egypt&                0.30&   	            H1N1 flu              &0.32\\
    H1N1 virus&  	           0.30&                Swine Flu             &0.26\\
    flu&	                   0.29&                swine flu Egypt       &0.24\\
    fever&	                   0.28&                News swine flu        &0.21\\
    swine&	                   0.27&                                      &    \\
    H1N1 flu&	               0.27&                                      &    \\
    Egypt swine flu&           0.13&                                      &    \\
    H1N1 in Egypt&   	       0.10&                                      &    \\    \cline{1-4}
    \multicolumn{4}{l}{NA: Not applicable} \\
    \multicolumn{4}{l}{\emph{p}$<$0.05} \\
  \end{tabular}}

In order to find the search queries with the highest correlation compared with the H1N1 morbidity trend data, Greedy Method was used to find the best combination of search queries. Meanwhile  we also considered the effects for  search queries  when  the  H1N1 morbidity data were shifted ahead or behind the Google search data for one or two weeks, so we must repeat the experiment using the different  H1N1 morbidity data.   We first calculated the correlation coefficients between the Google Trends data of each search query and the H1N1 morbidity  data of Egypt, and ranked them. Then the Greedy Method started with the search query which had the highest correlation coefficient, and added the other search queries  whose addition provided higher correlation coefficients for Google Flu Model. Repeating the above steps when the H1N1 morbidity data were shifted behind or ahead the Google search data for one or two weeks in Google Flu Model. Finally, we chose the  set of search queries with the highest correlation coefficient.

\section{Results}
\begin{figurehere}
  \centerline{\psfig{file=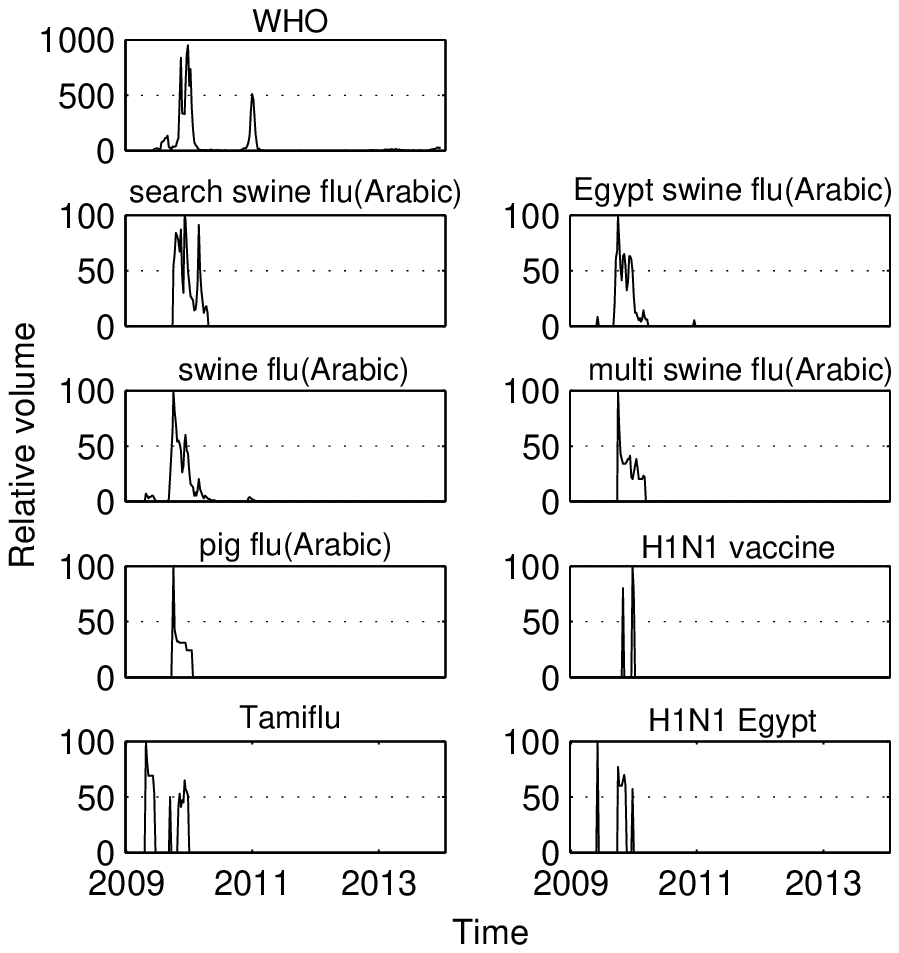,width=7cm}} 
  \vspace*{10pt}
  \fcaption{Time series plots of the H1N1 morbidity data of Egypt and Google Trends data.}
\end{figurehere}
\vspace*{13pt}

This study used the data from January 4, 2009 to December 31, 2013(261 weeks). First we chose 26 out of 50 search queries whose correlation coefficients were higher than 0 (Table 1). Then we got a set of 8 search queries with the highest correlation coefficient using the Greedy Method when the H1N1 morbidity data was shifted behind the Google search data for two weeks, they were \emph{search swine flu (Arabic)}, \emph{Egypt swine flu (Arabic)}, \emph{Swine flu (Arabic)}, \emph{multi swine flu (Arabic)}, \emph{pig flu (Arabic)}, \emph{H1N1 vaccine}, \emph{Tamiflu} and \emph{H1N1 Egypt} (Figure 1, 2). The correlation coefficients between the Google Trends for 8 queries and the H1N1 influenza surveillance data of Egypt ranged from 0.30 (\emph{p}$<$0.05) to 0.58 (\emph{p}$<$0.05) during the overall study period (Table 2). The highest correlation coefficient was \emph{search swine flu (Arabic)} of 0.58 (\emph{p}$<$0.05),
and the lowest was \emph{H1N1  Egypt} of 0.30 (\emph{p}$<$0.05).

\begin{figurehere}
  \vspace*{13pt}
  \centerline{\psfig{file=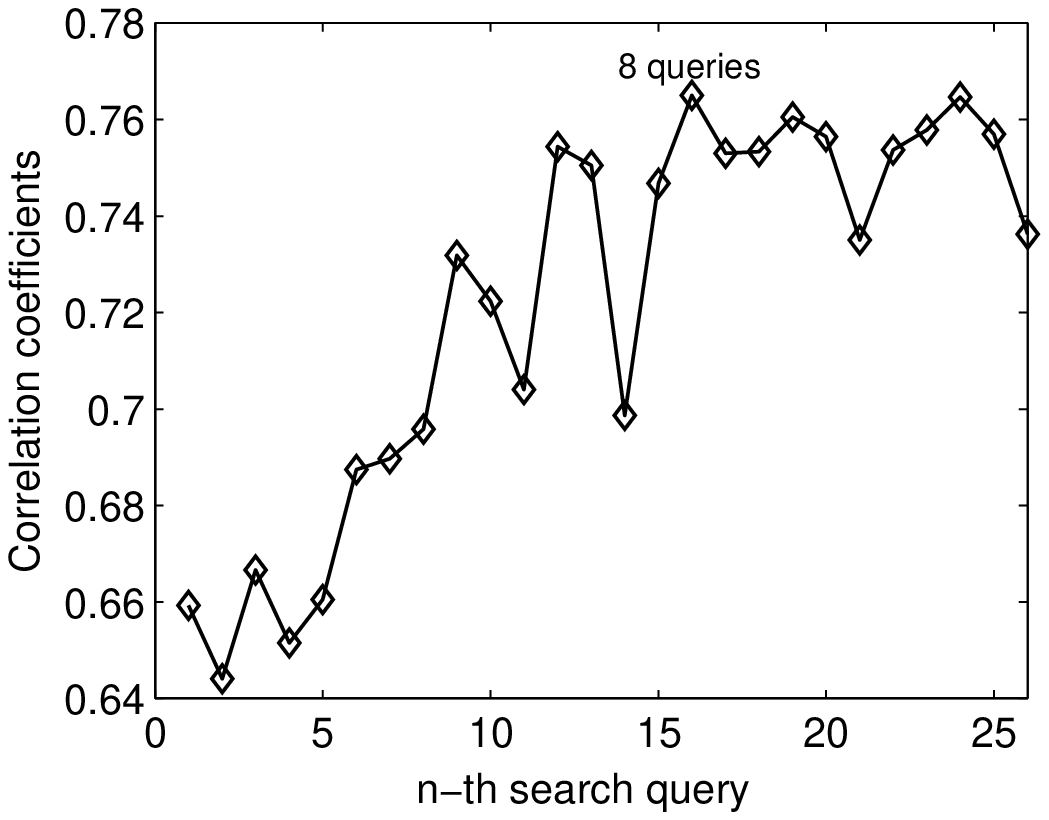,width=7cm}} 
  \vspace*{10pt}
  \fcaption{An evaluation of how many top-scoring queries to  fit the Google Flu Model when the H1N1 morbidity data of Egypt were shifted behind  the search query data for a  two-week.}
\end{figurehere}

\vglue13pt
\begin{table*}
  \caption{Pearson's correlation coefficients between the Google Trends data and the H1N1 data of Egypt  from 2009 to 2013.}
  \vglue-6pt
  \centerline{\small\baselineskip=13pt
  \begin{tabular}{c c c c c c c c c}\\
  \hline
  &\multicolumn{8}{c}{ \textbf{Google Trends}} \\  \cline{2-9}
  \textbf{Year}& \tabincell{c}{search \\swine flu}& \tabincell{c}{Egypt\\swine flu}& Swine flu& \tabincell{c}{multi \\swine flu}& pig flu& \tabincell{c}{H1N1 \\vaccine}& Tamiflu& \tabincell{c}{H1N1\\ Egypt} \\  \hline
  2009-2013&	0.58& 	0.53& 	0.49& 	0.49& 	0.48& 	0.43& 	0.37& 	0.30\\ \hline
  2009&	        0.66& 	0.53& 	0.45& 	0.40& 	0.39&	0.41& 	0.36& 	0.29\\ \hline
  2010&   	    0.29& 	0.70& 	0.69& 	0.65& 	0.88& 	0.51& 	NA&	    NA\\   \hline
  2011&	        NA&	    NA&	    NA&	    NA&	    NA& 	NA&	    NA&	    NA\\   \hline
  2012&	        NA&	    NA&	    NA&	    NA&	    NA& 	NA&	    NA&	    NA\\   \hline
  2013&	        NA&  	NA& 	NA& 	NA& 	NA& 	NA& 	NA&	    NA\\   \hline
  \multicolumn{9}{l}{NA: Not applicable} \\
  \multicolumn{9}{l}{\emph{p}$<$0.05} \\
  \end{tabular}}
\end{table*}

For annual correlation, the search query with the highest correlation coefficient  was \emph{pig flu (Arabic)}, 0.88 (\emph{p}$<$0.05), in 2010.  And \emph{Egypt swine flu} (\emph{Arabic} 0.70, \emph{p}$<$0.05) also showed strong correlation in 2010.  It was worth noting that the H1N1 epidemic was serious in 2009, the correlation coefficients of 8 key search queries were below 0.70. At the beginning of 2011, the H1N1 infections increased rapidly in Egypt, but the volume of 8 search queries were too low to be captured (Figure 1). And the volume  of 8 search queries were close to zero for the following two years (Figure 1).

The total  correlation coefficients of 8 search queries  increased when the H1N1 case data of Egypt  were shifted behind the Google Trends data with a one- or two-week (Table 3). This phenomenon also was found in 2009. When the epidemic was most serious in 2009, the highest correlation coefficient was only 0.81 (\emph{p}$<$0.05), and the other search queries ranged from 0.30 to 0.63 (\emph{p}$<$0.05) with a two-week lagging. For the other 4 years', we didn't find this except one search query, \emph{H1N1 vaccine}, in 2010, it's because the volume of search queries were too low to be captured (Figure 1).

\vglue13pt
\begin{table*}
  \caption{Pearson's correlation coefficients between the Google Trends data  and the time-shifting the H1N1 data of Egypt from 2009 to 2013.}
  \vglue-6pt
  \centerline{\small\baselineskip=13pt
  \begin{tabular}{clcccccccc}\\
  \hline
  \textbf{Year}&	\textbf{Dataset}& \tabincell{c}{search \\swine\\ flu} &\tabincell{c}{Egypt \\swine\\ flu}&\tabincell{c}{swine \\ flu}&\tabincell{c}{Multi\\swine\\ flu}	&\tabincell{c}{pig\\ flu}&\tabincell{c}{H1N1\\vaccine}&	Tamiflu &\tabincell{c}{H1N1 \\ Egypt}\\ \hline
        &WHO,2-week preceding&    0.40& 	0.35& 	0.34& 	0.48& 	0.38& 	0.36& 	0.22& 	0.07\\ \cline{2-10}
  2009  &WHO,1-week preceding&     0.48& 	0.42& 	0.40& 	0.49& 	0.44& 	0.48& 	0.29& 	0.17\\ \cline{2-10}
   |    &WHO,0-week lagging&      0.58& 	0.53& 	0.49& 	0.49& 	0.48& 	0.43& 	0.37& 	0.30\\ \cline{2-10}
  2013  &WHO,1-week lagging&      0.65& 	0.61& 	0.56& 	0.50& 	0.51& 	0.44& 	0.43& 	0.31\\ \cline{2-10}
        &WHO,2-week lagging&      0.70& 	0.65& 	0.61& 	0.53& 	0.53& 	0.50& 	0.45& 	0.36\\ \hline
        &WHO,2-week preceding&	  0.45& 	0.40& 	0.34& 	0.38& 	0.34& 	0.32& 	0.33& 	0.04\\ \cline{2-10}
        &WHO,1-week preceding&	  0.55& 	0.44& 	0.36& 	0.38& 	0.36& 	0.43& 	0.33& 	0.16\\ \cline{2-10}
  2009  &WHO,0-week lagging&	  0.66& 	0.53& 	0.45& 	0.40& 	0.39& 	0.41& 	0.36& 	0.29\\ \cline{2-10}
        &WHO,1-week lagging&	  0.75& 	0.61& 	0.53& 	0.47& 	0.43& 	0.35& 	0.40& 	0.27\\ \cline{2-10}
        &WHO,2-week lagging&	  0.81& 	0.63& 	0.57& 	0.53& 	0.48& 	0.51& 	0.37& 	0.30\\ \hline
        &WHO,2-week preceding&	  0.38& 	0.78& 	0.76& 	0.81& 	0.83& 	0.51& 	NA& 	NA\\ \cline{2-10}
	    &WHO,1-week preceding&	  0.37& 	0.82& 	0.78& 	0.76& 	0.92& 	0.67& 	NA& 	NA\\ \cline{2-10}
  2010  &WHO,0-week lagging&	  0.29& 	0.70& 	0.69& 	0.65& 	0.88& 	0.51& 	NA& 	NA\\ \cline{2-10}
	    &WHO,1-week lagging&	  0.20& 	0.65& 	0.60& 	0.41& 	0.68& 	0.68& 	NA& 	NA\\ \cline{2-10}
	    &WHO,2-week lagging&	  0.04& 	0.38& 	0.35& 	0.17& 	0.36& 	0.41& 	NA& 	NA\\ \hline
	    &WHO,2-week preceding&	  NA& 	NA& 	0.52& 	NA&  	NA& 	NA& 	NA& 	NA\\ \cline{2-10}
	    &WHO,1-week preceding&	  NA& 	NA& 	0.81& 	NA& 	NA& 	NA& 	NA& 	NA\\ \cline{2-10}
  2011  &WHO,0-week lagging&	  NA& 	NA& 	0.94& 	NA& 	NA& 	NA& 	NA& 	NA\\ \cline{2-10}
	    &WHO,1-week lagging&	  NA& 	NA& 	0.98& 	NA& 	NA& 	NA& 	NA& 	NA\\ \cline{2-10}
	    &WHO,2-week lagging&	  NA& 	NA& 	0.97& 	NA& 	NA& 	NA& 	NA& 	NA\\ \hline
	    &WHO,2-week preceding&	  NA& 	NA& 	NA& 	NA& 	NA& 	NA& 	NA& 	NA\\ \cline{2-10}
	    &WHO,1-week preceding&	  NA& 	NA&  	NA& 	NA& 	NA& 	NA& 	NA& 	NA\\ \cline{2-10}
  2012  &WHO,0-week lagging&	  NA& 	NA& 	NA& 	NA& 	NA& 	NA& 	NA& 	NA\\ \cline{2-10}
  	    &WHO,1-week lagging&	  NA& 	NA& 	NA& 	NA& 	NA& 	NA& 	NA& 	NA\\ \cline{2-10}
	    &WHO,2-week lagging&	  NA& 	NA& 	NA& 	NA& 	NA& 	NA& 	NA& 	NA\\ \hline
	    &WHO,2-week preceding&	  NA& 	NA& 	NA& 	NA& 	NA& 	NA& 	NA& 	NA\\ \cline{2-10}
  	    &WHO,1-week preceding&	  NA& 	NA& 	NA& 	NA& 	NA& 	NA& 	NA& 	NA\\ \cline{2-10}
  2013  &WHO,0-week lagging&	  NA& 	NA& 	NA& 	NA& 	NA& 	NA& 	NA& 	NA\\ \cline{2-10}
	    &WHO,1-week lagging&	  NA& 	NA& 	NA& 	NA& 	NA& 	NA& 	NA& 	NA\\ \cline{2-10}
	    &WHO,2-week lagging&	  NA& 	NA& 	NA& 	NA& 	NA& 	NA& 	NA& 	NA\\ \hline
  \multicolumn{10}{l}{NA: Not applicable} \\
  \multicolumn{10}{l}{\emph{p}$<$0.05} \\
  \end{tabular}}
\end{table*}

For the Google Flu Model, we achieved a combination of 8 search queries with the highest correlation coefficient using Greedy Method, and the others were in table 4 when the H1N1 morbidity data were shifted behind or ahead the Google search data. We found that the highest correlation coefficient  was 0.76 (\emph{p}$<$0.05, the H1N1 data 2-week lagging, Figure 2).  And the annual correlation coefficients were 0.83, 0.77, -0.76, -0.37, -0.29 (\emph{p}$<$0.05) from 2009 to 2013, respectively.

\vglue13pt
\begin{table*}
  \caption{Pearson's correlation coefficients between the Google Trends data and the H1N1 data of Egypt  from 2009 to 2013.}
  \vglue-6pt
  \centerline{\small\baselineskip=13pt
  \begin{tabular}{c c c c c c}\\
  \hline
  Dataset&2-week preceding&1-week preceding&0-week lagging&1-week lagging&2-week lagging\\ \hline
  &0.61&0.65&0.71&0.74&0.76\\ \hline
  \multicolumn{6}{l}{\emph{p}$<$0.05} \\
  \end{tabular}}
\end{table*}

\section{Discussion}
For the influenza  related search query,  \emph{Tamiflu}, its first peak of the search volume occurred   at the end of April, 2009 (Figure 1), but there was little infected people (the H1N1 data published by WHO in Figure 1). It was widely reported when H1N1 broke out in Mexico in April, 2009\up{6}. This attracted a lot of people to search the treatment, precautionary
measures, and other information about H1N1 on the Internet.  In the early stages of infectious disease, the Google Trends data would have a peak, but in fact the rush had nothing to do
with the H1N1 morbidity trends. And we thought that phenomenon was related with media reports.

The search volume of queries was large at 2009, but the amount were small in the following years (Figure 1). For one thing, people paid less attention on H1N1 when epidemic was under
control. For another, people had known the symptoms and prevention methods after one year of epidemic, so they didn't need to search on the website. These factors had a big influence on
the prediction model like Google Flu Trends.  We could see it from model estimates, especially  when the epidemic broke out again at 2011. So people's behavior on the Internet can not
explain this condition whether in the early or late of outbreak. This point was also useful to design a total influenza  monitoring system. And  because of the changeable of people's
search behavior,  the algorithms of  Google Flu Model  need to be revised constantly\up{12,17}\ .

From the correlation coefficients between the H1N1 morbidity data of Egypt and the Google Trends data (ranged from 0.30 to 0.58, p$<$0.05), we found that it was smaller than the previous
related researches\up{12,17,21}. When search volume was large at 2009, the correlation coefficient of 8 search queries did not show strong correlation (ranged from 0.29 to 0.66). But when
time was a two-week lagging, the correlation coefficients of 8 search queries increased in 2009(correlation coefficient ranged from 0.30 to 0.81). This was similar with other researches\up{5,13}. And some studies shown that using search data to predict flu trends can be 1-2 weeks earlier than traditional influenza surveillance system\up{12,13,15,16}. It was indicated that there was an increasing correlation coefficient between the search engine data and  the H1N1 Morbidity data with a two-week lagging. Then this had obvious practical implications to Google Flu Model.

When we used Google model to predict, the result of correlation coefficient was improved, but remained inaccurate (Figure 3). We found that the number of affected people was lower than
zero in September 2009 (Figure 3).  And there were two peaks from December 2009 to January 2010, when the model estimates raised and decreased, rapidly, it was consistent with the real
situation. However  the maximum and minimum values were not identical to the real values, and the other papers also found it\up{12,17,22}. The  model estimates showed that the number of H1N1 infection would surge in March 2010 (Figure 3). But in fact, this was a false alarm, and this did not help to play the role in early warning. The epidemic in Egypt peaked at 2011,
but the model estimates didn't show the same situation, and showed larger discrepancies. Though we picked dozens of search queries (Table 1), most search queries'  volume  were small, or
correlation coefficients were low. From table 2 and 3, we could see that the volume of 8 search queries were close to zero from 2010 to 2013, and that was why the model estimates were not
good. It also showed that using search engines to predict the flu trends would have the same problem for the other developing countries in Africa.

\begin{figurehere}
  \vspace*{13pt}
  \centerline{\psfig{file=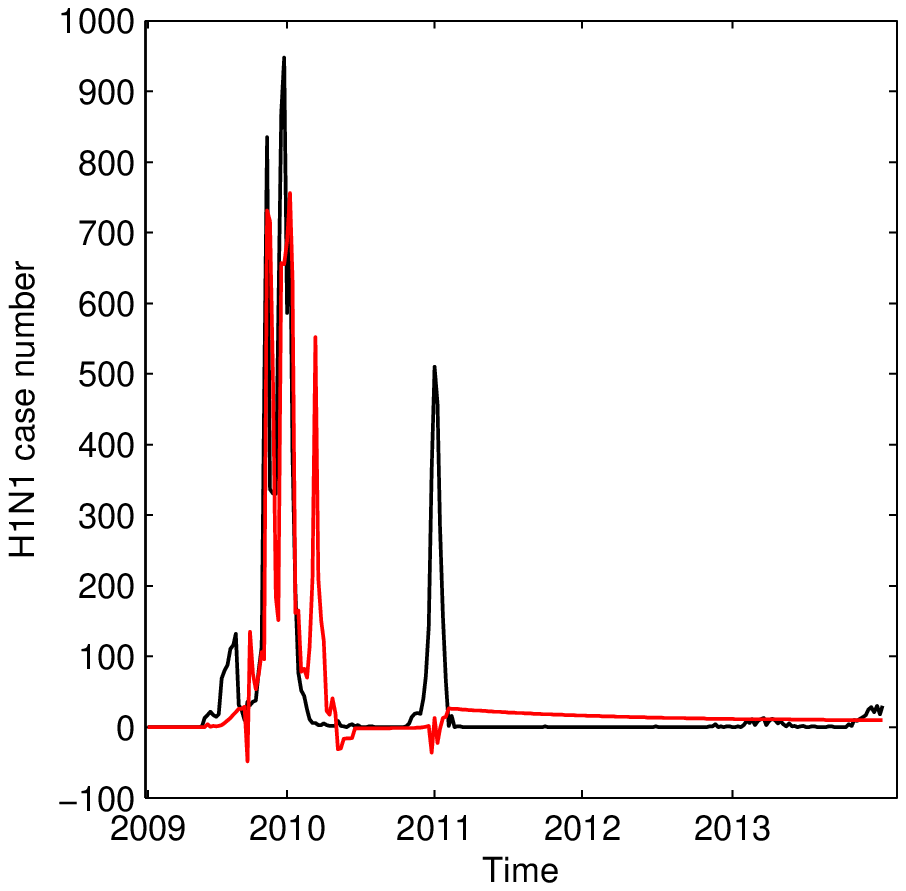,width=7cm}} 
  \vspace*{10pt}
  \fcaption{A comparison of Google Flu Model estimates for Egypt(red) against the WHO-reported the number of H1N1 infection(black), and what were fitted for the Google Flu Model were  the H1N1 influenza surveillance data, shifted behind the Google Trends data for  two weeks. (Correlation coefficient was 0.76(\emph{p}$<$0.05))}
\end{figurehere}
\vspace*{13pt}

There are some limitations in this study. First, we just calculated the correlation coefficient but didn't search the causality. Second, there was a limitation in searching on website
because of unbalanced economics and religious culture in Egypt. Third, this research was finished after many years of epidemic, and selected some search queries related to H1N1, it maybe miss some search queries related to H1N1. Fourth, the WHO data would miss or lack some important information. At last, the Google Trends data was normalized and changeable, it may lead to the change in correlation coefficient.

In conclusion,  there was a positive correlation between the Google Trends data and the H1N1 morbidity data of Egypt, but the correlation were relatively lower than previous researches.
Though the correlation coefficients between the model estimates data and the H1N1 morbidity data was more than 0.70 (\emph{p}$<$ 0.05), the model estimates   were not accurate. Because the  people's behavior on the Internet changed all the time, and the model  had to  tweak algorithms constantly.  There should be more researches using search queries about predicting the flu trends in the future, not only in the developed countries, but also in the developing countries.

\section*{References}


\end{multicols}

\begin{thebibliography}{000}
\bibitem{1}
Wikipedia(2014) Influenza. Available:http://en.wiki- pedia.org/wiki/Flu. Accessed June 20th, 2014.

\bibitem{2}
Frenk J, Gomez Dantes O, ``Globalization And The Challenges To Health Systems,'' {\it Health Affairs},{\bf21}, 160--165 (2002).

\bibitem{3}
Irvin CB, Nouhan PP, Rice K, ``Syndromic analysis of computerized emergency department patients¡¯ chief complaints:An opportunity for bioterrorism and influenza surveillance'' {\it Annals of Emergency Medicine},{\bf41}, 447-452, (2003).

\bibitem{4}
Eysenbach G, ``Infodemiology: tracking flu-related searches on the web for syndromic surveillance,'' {\it AMIA Annu Symp Proc}, 244--248, (2006).

\bibitem{5}
Cho S, Sohn CH, Jo MW, Shin S-Y, Lee JH, et al, ``Correlation between National Influenza Surveillance Data and Google Trends in South Korea,'' {\it PLoS one},{\bf8}(12), e81422 (2013).


\bibitem{6}
Wikipedia (2014) Influenza A virus subtype H1N1, Available:http://en.wikipedia.org/wiki/Influenza\_A \_virus\_subtype\_H1N1. Accessed June 20th, 2014.

\bibitem{7}
World Health Organization(Cases Reports of Influenza Surveillance all over world) Avaailable: http://gamapserver.who.int/gareports/Default.aspx? ReportNo=7. Accessed June 20th, 2014.

\bibitem{8}
Varney S.M, Hirshon J.M, ``Update on public health surveillance in emergency departments,'' {\it Emerg Med Clin North Am},{\bf24},1035--1052 (2006).

\bibitem{9}
Henning KJ, ``What is syndromic surveillance?'' {\it MMWR Morb Mortal Wkly Rep 53 Suppl}, 5--11 (2004).

\bibitem{10}
Yang AC, Huang NE, Peng C.K, Tsai S.J, ``Do seasons have an influence on the incidence of depression? The use of an internet search engine query data as a proxy of human affect,'' {\it PloS one} {\bf5}, e13728 (2010).

\bibitem{11}
Polgreen PM, Chen Y, Pennock DM, Nelson FD, ``Using Internet Searches for Influenza Surveillance,'' {\it Clinical infectious diseases: an official publication of the Infectious Diseases Society of America}, {\bf47}, 1443--1448  (2008).

\bibitem{12}
 Ginsberg J, Mohebbi MH, Patel RS, Brammer L, Smolinski MS, et al, ``Detecting influenza epidemics using search engine query data,'' {\it Nature}, {\bf457}, 1012--1014 (2009).

\bibitem{13}
Pelat C, Turbelin C, Bar-Hen A, Flahault A, Valleron A, ``More Diseases Tracked by Using Google Trends,'' {\it Emerg Infect Dis}, {\bf15}(8), 1327--8, (2009).

\bibitem{14}
Google (2014) Google Flu Trends. Available: http://www.google.org/flutrends/intl/en\_us/. Accessed June 20th, 2014.

\bibitem{15}
Ortiz JR, Zhou H, Shay DK, Neuzil KM, Fowlkes AL, et al, ``Monitoring influenza activity in the United States: a comparison of traditional surveillance systems with Google Flu Trends,''  {\it PloS one}, {\bf6}, e18687,(2011).

\bibitem{16}
Valdivia A, Lopez Alcalde J, Vicente M, Pichiule M, Ruiz M, et al, ``Monitoring influenza activity in Europe with Google Flu Trends: comparison with the findings of sentinel physician networks  results for 2009--10,'' {\it Euro surveillance}, {\bf15},(2010).

\bibitem{17}
Cook S, Conrad C, Fowlkes AL, Mohebbi MH, ``Assessing Google flu trends performance in the United States during the 2009 influenza virus A(H1N1) pandemic,'' {\it PloS one}, {\bf6}, e23610, (2011).

\bibitem{18}
Wilson N, Mason K, Tobias M, Peacey M, Huang QS, et al, ``Interpreting Google flu trends data for pandemic H1N1 influenza: the New Zealand experience,'' {\it Euro surveillance: bulletin europeen surles maladies transmissibles= European communicable disease bulletin 14}, (2009).

\bibitem{19}
World Health Organization Flunet. Available: http://www.who.int/influenza/gisrs\_laboratory/en/. Accessed June 20th, 2014.

\bibitem{20}
Kang M, Zhong H, He J, Rutherford S, Yang F, ``Using Google Trends for Influenza Surveillance in South China,'' {\it PLoS ONE}, {\bf8}1, e55205, (2013).

\bibitem{21}
Olson DR, Konty KJ, Paladini M, Viboud C, Simonsen L, ``Reassessing Google Flu Trends Data for Detection of Seasonal and Pandemic Influenza: A Comparative Epidemiological Study at Three Geographic Scales,'' {\it PLoS Comput Biol}, {\bf9}, 10, e1003256, (2013).

\bibitem{22}Butler D, ``When Google got flu wrong,'' {\it Nature}, {\bf494}, 7436, 155--6, (2013).


\label{\labart-LastPage}
\end{thebibliography}
\end{document}